\documentstyle[aps,prd,epsf]{revtex}

\newcommand{\be}{\begin{equation}}
\newcommand{\ee}{\end{equation}}
\newcommand{\ben}{\begin{eqnarray}
\displaystyle}
\newcommand{\een}{\end{eqnarray}}

\newcommand{\la}{{\lambda}}

\newcommand{\de}{{\delta}}

\newcommand{\cH}{{\cal H}}

\newcommand{\cO}{{\cal O}}

\newcommand{\cS}{{\cal S}}

\newcommand{\cU}{{\cal U}}

\newcommand{\p}{\partial}
\newcommand{\na}{\nabla}

\newcommand{\Liek}{{\cal L}_{k^{\mu}}}

\newcommand{\tg}{{\tilde g}}

\newcommand{\hpsi}{{\hat \psi}}

\newcommand{\hro}{{\hat \rho}}

\newcommand{\ep}{\epsilon}

\newcommand{\ga}{\gamma}

\begin{document}

\title{Classification of Static Charged Black
Holes
in Higher Dimensions}

\author{Marek Rogatko}

\address{Institute of Physics \protect \\
Maria Curie-Sklodowska University \protect \\
20-031 Lublin, pl.Marii Curie-Sklodowskiej 1, Poland \protect \\
rogat@tytan.umcs.lublin.pl \protect \\
rogat@kft.umcs.lublin.pl}

\date{\today}

\maketitle
\begin{abstract}
The uniqueness theorem for static charged higher dimensional black hole
containing an asymptotically flat spacelike hypersurface with compact interior
and with both degenerate and non-degenerate components of event horizon
is proposed. By studies of the near-horizon geometry of degenerate horizons
one was able to eliminate the previous restriction concerning the inequality
fulfilled by the charges of the adequate components of the aforementioned horizons.
\end{abstract}


\baselineskip=18pt
\section{Introduction}
The main aim of the uniqueness theorems of black holes in general relativity
is to show that the static electrovac black hole spacetime are described by
Reissner-Nordstr\"om (RN) spacetime whereas the circular one is diffeomorphic to Kerr-Newman (KN)
spacetime. Israel \cite{isr} was the first to raise the problem of classification
of non-singular black hole solutions.
M\"uller zum Hagen {\it et al.} \cite{mil73} and Robinson \cite{rob77},
were able to weaken Isreal's assumptions while
the most complete results were proposed in Refs.\cite{bun87,ru,ma1,he1,he93}.
In Ref.\cite{chr99a}
the classification of static vacuum black hole solutions was finished 
in \cite{chr99a} (see also the latest refinement of the proof \cite{chr06}), where
the condition of non-degeneracy of the 
event horizon was removed. In Ref.\cite{chr99b} it was shown that for the static
electro-vacuum black holes all degenerate components of 
the event horizon should have charges of the same signs. Recently, by study of 
the near-horizon geometry this last restriction was removed \cite{chr05}.
\par
The uniqueness black hole theorem for stationary axisymmetric
spacetime was elaborated in Refs.\cite{stat}, while the complete proof was found by
Mazur \cite{maz} and Bunting \cite{bun}
(see for a review of the uniqueness of black hole
solutions story see \cite{book} and references therein).
\par
The unification attempts such as M/string theory
caused the  
resurgence of works devoted to the mathematical aspects of the low-energy string theory
black holes as well as higher dimensional ones. Namely,
the staticity theorem for Einstein-Maxwell axion dilaton (EMAD) gravity
was studied in Ref.\cite{sta}. Then,
the uniqueness of the black hole solutions in dilaton gravity was  proved in 
works
\cite{mas93,gur95,mar01}, while the uniqueness of the static
dilaton $U(1)^2$
black holes being the solution of $N = 4,~ d = 4$ supergravity
was provided in \cite{rog99}. The extension of the proof to the theory
to allow for the inclusion of 
$U(1)^{N}$ static dilaton black holes was established in Ref.\cite{rog02}.
\par
On the other hand, the uniqueness theorem for
$n$-dimensional black hole, both in vacuum and
charged case were proposed \cite{gib03,gib02a,gib02b,kod04}. The complete 
classification of $n$-dimensional
charged black holes having both degenerate and non-degenerate
components of event horizon was provided in Ref.\cite{rog03}.
In Ref.\cite{rog04}, taking into account the both {\it electric} and {\it magnetic} components of 
$(n-2)$-gauge form $F_{\mu_{1} \dots \mu_{n-2}}$ the uniqueness
of static higher dimensional {\it electrically} and {\it magnetically} charged
black hole containing an asymptotically flat hypersurface with compact interior
and non-degenerate components of the event horizon was proved. On the other hand, the staticity theorem
for generalized Einstein-Maxwell (EM) system was discussed in \cite{rog05}.      
\par
Proving the uniqueness theorem for stationary 
$n$-dimensional black holes is much more complicated.
It turned out that generalization of Kerr metric to arbitrary $n$-dimensions proposed by
Myers-Perry \cite{mye86} is not unique. The counterexample showing that a five-dimensional
rotating black hole ring solution with the same angular momentum and mass
but the horizon of which was homeomorphic to $S^{2} \times S^{1}$ was presented in
\cite{emp02} (see also Ref.\cite{emp04}). 
The existence of black rings is consistent with the generalization of Hawking's
theorem \cite{haw73}. It turned out that \cite{gal05} cross sections of event horizons and
outer horizons are of positive Yamabe type, i.e., they admit metrics of positive scalar curvature.
In Ref.\cite{hel06} the possible geometric types of event horizons for five and higher 
dimensional spacetimes were discussed.
Recently, it has been established that
Myers-Perry solution is the unique black hole in five-dimensions in the class of 
spherical topology
and three commuting Killing vectors \cite{mor04}.
The uniqueness theorem for self-gravitating nonlinear $\sigma$-models
in higher dimensional spacetime was obtained in \cite{rog02a}.
\par
Our paper will be devoted to the uniqueness of static higher dimensional charged black holes
in generalized EM system described by $(n-2)$-gauge form $F_{\mu_{1} \dots \mu_{n-2}}$
containing an asymptotically flat spacelike hypersurface with compact interior
with both degenerate and non-degenerate components of the event horizon. Studies of the near-horizon
geometry of degenerate black hole enables one to eliminate the assumption concerning the charge of the
adequate components of the horizons, i.e., to eliminate the last restriction in the complete
classification of charged static $n$-dimensional black holes described in \cite{rog03}.

\section{Higher dimensional generalized Einstein-Maxwell system}
The action of 
the generalized Maxwell $(n-2)$-gauge
form $F_{\mu_{1} \dots \mu_{n-2}}$ in $n$-dimensional spacetime we consider here is given by
\be
I = \int d^n x \sqrt{-{g}} \bigg[ {}^{(n)}R - 
F_{(n-2)}^2
\bigg],
\label{act}
\ee
where ${g}_{\mu \nu}$ is $n$-dimensional metric tensor,
$F_{(n-2)} = dA_{(n-3)} $ is $(n-2)$-gauge form field.
The energy momentum tensor of the $(n-2)$-gauge form 
$T_{\mu \nu} = - {\de S \over \sqrt{- g}\de g^{\mu \nu}}$ yields
\be
T_{\mu \nu} = (n -2) F_{\mu i_{2} \dots i_{n-2}}
F_{\nu}{}{}^{i_{2} \dots i_{n-2}} - {g_{\mu \nu} \over 2}F_{(n-2)}^2.
\ee
We shall confine our attention to the asymptotically flat spacetime. Thus, the spacetime with 
contain a data set $(\Sigma_{end},~g_{ij},~K_{ij})$ with gauge field determined by $(n - 2)$
gauge form field $F_{i_{1} \dots i_{n-2}}$ for which $\Sigma_{end}$ is
diffeomorphic to ${\bf R}^{n-1}$ minus a ball.
The asymptotical conditions imply
\ben
\mid g_{ij} - \delta_{ij} \mid +~ r \mid \p_{a} g_{ij} \mid + \dots
+r^{m} \mid \p_{a_{1} \dots a_{m}} g_{ij} \mid + r \mid K_{ij}\mid + \dots
+r^{m} \mid \p_{a_{1} \dots a_{m}} K_{ij} \mid \le {\cal O}\bigg( {1 \over r} \bigg), \\ 
\mid F_{i_{1} \dots i_{n-2}} \mid +~ 
r \mid \p_{a} F_{i_{1} \dots i_{n-2}} \mid + \dots +
r^{m} \mid \p_{a_{1} \dots a_{m}}F_{i_{1} \dots i_{n-2}} \mid
\le {\cal O}\bigg( {1 \over r^{2}} \bigg).
\een
First, as in Ref.\cite{rog03} we introduce the
generalization of the definition of electric components for the ordinary
Maxwell fields. One can define {\it electric} $(n-3)$-form by the expression as
follows:
\be
E_{i_{1} \dots i_{n-3}} = F_{i_{1} \dots i_{n-2}} k^{i_{n-2}}.
\label{elec}
\ee
On the other hand {\it magnetic} $1$-form may be written as
\be
B_{k} = {1 \over \sqrt{2 (n -2)!}}~ \ep_{k \mu i_{1} \dots i_{n-2}}
F^{i_{1} \dots i_{n-2}} k^{\mu}.
\label{mag}
\ee
We introduce also the rotation $(n-3)$-form of the stationary
Killing vector field $k_{\mu}$ in the form of
\be
\omega_{j_{1} \dots j_{n-3}} = {(n-2) \over \sqrt{2 (n -2)!}}~
\ep_{j_{1} \dots j_{n-3} \mu \nu \gamma} k^{\mu} \na^{\nu} k^{\gamma}.
\label{tw}
\ee
Using the definition (\ref{tw}) and definition of {\it electric} and
{\it magnetic} forms and 
having in mind
equations of motion for $(n-2)$-gauge form fields
one achieves at Eqs. of motion for
{\it magnetic} $1$-form $B_{k}$
\be
\na^{a}\bigg( {B_{a} \over N }\bigg) = {E^{i_{1} \dots i_{n-3}}~
\omega_{i_{1} \dots i_{n-3}} \over N^2},
\label{bbb}
\ee
and similarly for {\it electric} $(n-3)$-form $E_{i_{1} \dots i_{n-3}}$
\be
\na^{j_{1}} \bigg( {E_{j_{1} j_{2}  \dots j_{n-3}}\over N}\bigg) = - {B^{a}~
\omega_{a j_{2} \dots j_{n-2}}\over N^2},
\label{eee}
\ee
where we have denoted $N = k_{\mu}k^{\mu}$.\\
If one applies the relation valid for the Killing vector fields $\na_{\alpha} \na_{\beta} \xi_{\ga}
= - R_{\alpha \beta \ga}{}{}{}^{\delta} ~\xi_{\delta}$ and definitions (\ref{elec}) and (\ref{mag})
to equation (\ref{tw}), it can be verified that
\be
\omega_{[j_{1} \dots j_{n-3}; b]} = - \alpha~E_{[j_{1} \dots j_{n-3}}~B_{b]},
\ee
where $\alpha = {\sqrt{2} (n - 2 )^{3/2} \over \sqrt{(n - 3)!}}$.\\
For we look for the static metric the right-hand sides of Eqs. (\ref{bbb}) and (\ref{eee}) 
are equal to zero and we have
\be
\ep_{j_{1} \dots j_{n-3} ~j_{n-2}}{}{}{}{}^{i_{1} \dots i_{n-3} k}
E_{i_{1} \dots i_{n-3}}~B_{k} = 0.
\ee
If we choose {\it electric} $(n-3)$-form as $E_{i_{1} \dots i_{n-3}} =
\delta^{0}_{i_{1}}~\delta^{0}_{i_{2}} \dots \delta^{m}_{i_{n-3}}~\na_{m} \phi$
and {\it magnetic} $1$-form $B_{k} = \na_{k} \psi$ one can conclude 
that $\psi = c~\phi$. Using equations of motion for magnetic $B_{k}$ $1$-form in the static case, i.e.,
\be
\na^{a}\bigg( {B_{a} \over N }\bigg) = 0,
\ee
and multiplying it by $\delta^{0}_{i_{1}}~\delta^{0}_{i_{2}} \dots \delta^{m}_{i_{n-3}}$
it then follows directly
that $c = const$. Thus, by virtue of
the hypothesis of staticity and by means of choosing the special components of 
$E_{i_{1} \dots i_{n-3}}$, the {\it magnetic} field can be made vanish by a duality rotation. Namely,
suppose that $c$ is constant on each connected component of the set 
$\Omega$, 
where $E_{i_{1} \dots i_{n-3}} \neq 0$
and suppose moreover that $\Omega_{0}$ be any connected component of the above set.
By performing a duality rotation we can obtain
$\psi = 0$ in $\Omega_{0}$. 
Because of the fact that $\Omega_{0}$ is open set, having in mind equations of
motion and the unique 
continuation theorem of Ref.\cite{aro57} one has that $\psi = 0$ and therefore
$B_{k} = 0$.
\par
From now on we shall consider only the {\it electric} component of 
$F_{i_{1} \dots i_{n-2}}$. As in four-dimensional case the idea of the proof will be to show 
that degenerate components of the event horizon are only admissible for
Majumdar-Papapetrou spacetimes. We prove that the metric of static degenerate horizon is spherical.
Then, one can use a conformal transformation to the electric potential, normalized in such a case that
it tends to zero at infinity, to prove that this geometry is possible to occur
with $\phi = \pm 1$ on a component of the horizon iff the metric is of Majumdar-Papapetrou type.
Then this problem reduces to the one where $\mid \phi \mid$ is strictly bounded away from one.
It was shown that it leads to the Reissner-Nordstr\"om geometry \cite{chr99b,rog03}.
\par
To begin with
we shall introduce a Gaussian null coordinates near the event horizon of black hole as in \cite{fri99,rea03}. Namely,
let $\Sigma$ be a Cauchy hypersurface for the exterior region of the black hole. Let us introduce 
local coordinates $x^{a}$, where $a = 1, \dots ,n-2$
on an event horizon. Consider a point $p(x^{a})$ on $\cH$ through which
a future directed geodesic with tangent vector $k_{a}$ will pass. The integral curves of vector field $k_{a}$
are the null generators of $\cH^{+}$. On a sufficiently small neighbourhood $\cS$ of a point
$p$, let $\theta : \cS \rightarrow \cH^{+}$
be the map which takes $\{p, v \}$ into point of $\cH^{+}$ lying at parameter
value $v$ along the integral curve of $k^{a}$ starting at point $p$. This procedure defines coordinates on a neighbourhood
$\cU \subset \cH^{+}$ with $k_{a} = \bigg( \p /\p v \bigg)_{a}$. 
At each point of $\cH^{+}$ let $l^{a}$ be the unique null vector
satisfying the conditions $ l^{a}k_{a} = 1 $ and $l^{a} X_{a} = 0$, for all 
$X_{a}$ tangent to the surface of constant $v$.
Let us consider further the map which takes $\{p, r \}$ into the point of manifold lying at
affine parameter $r$ along the null geodesic from $p \in \cU$ with the tangent vector $l^{a}$.
We have extended the functions $v,~ x^{a}$ by requiring that their values to be constant along each null
geodesics determined by $l^{a} = \bigg( \p /\p r \bigg)^{a}$. On the event horizon $\cH^{+}$
one has $\Liek l^{a} = 0$, moreover $k^a$ is a Killing vector field and hence geodesics 
are mapped to geodesics under the flow of the vector $k_{a}$. Since, the vector $l^a$ is everywhere tangent to null 
geodesics we have that $g_{rr} = 0$.
Because of the independence of $r$ coordinate, the metric
functions $g_{r v} = 1, ~ g_{r a} = 0$. The fact that $\p /\p x^{a}$ is tangent to the surface
of constant $v$ and hence orthogonal to $k_{a}$ at $r = 0$, then $g_{v a} = r h_{a}(r, x^a)$ for some
function $h_{a}$ independent of the $u$ coordinate. Hence, there exists smooth function 
$\varphi \mid_{\cH^{+}} = \p g_{vv} /\p r \mid_{r = 0}$ the spacetime metric implies
\be
ds^2 = r \varphi~ dv^2 + 2 dv dr + 2 r h_{a}~ dx^{a} dv + h_{ij}~ dx^{i} dx^{j},
\label{met}
\ee
where $ h_{ij}$ is a metric tensor on $(n-2)$-dimensional spacetime.\\
The location of the event horizon is given by the relation $r = 0$.
The surface gravity equals to $\kappa = - \p_{r} (r \varphi)$. Thus, for the degenerate horizon 
the surface gravity is zero so it implies the fact that
$\varphi = A(r, x^{a})~ r$. The form of the metric (\ref{met}) guarantees us the existence
of a regular near-horizon geometry which is defined by the limit $r \rightarrow \ep~r$ and
$v \rightarrow v/\ep$. Consequently, when $\ep \rightarrow 0$ the metric (\ref{met}) tends 
to the metric which can be expressed as
\be
ds_{(0)}^2 = r^2 A_{(0)}dv^2 + 2 dv dr + 2 r h_{(0) a}dx^{a}dv + h_{(0)ab}dx^{a} dx^{b},
\ee
with the additional conditions as follows:
\be
\p_{r}A_{(0)} = \p_{r}h_{(0) a} = \p_{r} h_{(0)ab} = \p_{v}A_{(0)} = \p_{v}h_{(0) a} = \p_{v} h_{(0)ab} = 0,
\ee
where we have denoted $A_{(0)} = A \mid_{r=0},~ h_{(0) a} = h_{a} \mid_{r=0}$ and $h_{(0)ab} = h_{ab} \mid_{r=0}.$
So the line element
$ds_{(0)}^2$ encompasses information about the behaviour of $h_{ab},~h_{a}$ and $A$ at the event horizon
$\cH$. Having in mind the form of the metric near horizon, after tedious calculations it can be shown that
$n$-dimensional generalized Einstein-Maxwell equations of motion yield
\be
{1 \over 2}h_{(0)a}h_{(0)b} - {}^{(n-2)} \na_{(a}h_{(0)b)} =
- {}^{(n-2)}R_{(0)ab} + T_{ab} + {T \over 2 - n}~h_{(0)ab},
\label{rr}
\ee
where $R_{(0)ab}$ is the Ricci tensor of metric $h_{(0)ab}$ and ${}^{(n-2)} \na_{a}$ is a derivative
with respect to this metric. As was shown in Ref.\cite{chr06,chr05} the staticity of $g_{\mu \nu}$ implies
the staticity of $g_{(0)\mu \nu}$, which in turns leads to the fact that at the event horizon
$h_{a}$ is a gradient, namely $h_{a}~dx^{a} = d\la.$ It helps us to rewrite Eq.(\ref{rr}) as
\be
{}^{(n-2)}\na_{a} {}^{(n-2)}\na_{b}~\la - {1\over 2} {}^{(n-2)}\na_{a} \la
{}^{(n-2)}\na_{b} \la = - {}^{(n-2)}R_{(0)ab} + 2 (n - 3)^2 ~ \mid \na \phi \mid^2 h_{(0)ab}.
\label{lam}
\ee
Further, we introduce $\psi = e^{- \la/2}$ so one finally gets the following:
\be
{}^{(n-2)}\na_{a} {}^{(n-2)}\na_{b}~\psi = {\psi \over 2}~ {}^{(n-2)}R_{(0)ab} - (n - 3)^2 ~ 
\mid \na \phi \mid^2 ~\psi~h_{(0)ab}.
\label{abp}
\ee
Consistently with the above 
the trace of the expression (\ref{abp}) may be determined by
\be
{{}^{(n-2)}\na}^2~ \psi = {\psi \over 2}~ {}^{(n-2)}R_{(0)} - (n - 2)(n - 3)^2~\mid \na \phi \mid^2 ~\psi.
\label{ppp}
\ee
After taking the ${{}^{(n-2)}\na}^{b}$ derivative of equation (\ref{abp}) and using
the relation (\ref{ppp}), one gets
\be
{}^{(n-2)}\na \bigg( \psi^{3}~{}^{(n-2)}R - {4 \over 3} B \mid \na \phi \mid^2~\psi^3 \bigg) = 0,
\ee
where we have denoted $B = (n - 3)[(n - 3)^{2} + {1 \over 2}(n - 2)]$. This is sufficient to establish that
the Ricci scalar curvature may be written in the form
\be
{}^{(n-2)}R_{(0)} = {3 c_{1} + 4 B \mid \na \phi \mid^2~\psi^{3} \over 3 \psi^{3}},
\ee
where $c_{1}$ is a constant.\\
In the light of what has been shown before, it can be noted that
one can rewrite Eq.(\ref{ppp}) and (\ref{abp}) as follows:
\be
{{}^{(n-2)}\na}^2~ \psi = {3 c_{1} + 2(n - 3)[(n - 2) - n(n - 3)]~\mid \na \phi \mid^2~ \psi^3 \over 6 \psi^2},
\label{pp1}
\ee
and
\be
{}^{(n-2)}\na_{a} {}^{(n-2)}\na_{b}~\psi = \bigg[
{3 c_{1} + 4(n - 3)\big[ (n - 3)(n - 6) + 2(n - 2) \big] \mid \na \phi \mid^2~\psi^3
\over 12 \psi^2} \bigg] h_{(0)ab}.
\label{pp2}
\ee
One can also find that
\be
{}^{(n-2)}\na^{m} \psi {}^{(n-2)}\na_{m} \psi = c_{2} - {c_{1} \over 2 \psi} +
{1 \over 6}~ K~ \mid \na \phi \mid^2~ \psi^2,
\ee
where $K = 2(n - 3)[(5 - n) + (n - 3)^2]/6$ and $c_{2}$ is some constant.\\
Now
we shall take up the question about the critical set of $\psi$.
We denote by  $a = \inf \psi$ and $b = \sup \psi$. Suppose further, that $a \neq b$. 
Because of the fact that $\psi$ is smooth and positive we have $0 < a \leq b < \infty$.
By virtue of the fact that for any point $p$ such that $\psi(p) = b$, one has that 
${}^{(n-2)}\na^{m} \psi (p) = 0$ and
${}^{(n-2)}\na^{m} \psi {}^{(n-2)}\na_{m} \psi (p) < 0$ examination of the relation (\ref{pp1})
reveals the following:
\be
c_{1} < - {2 (n - 3) \over 3} \bigg[ (n - 2) - n (n - 3) \bigg]~\mid \na \phi \mid^2~b^3,
\label{c1}
\ee
for $\na \phi \neq 0$. On further simplification, Eq.(\ref{pp2}) can be brought to 
the form revealing the condition for $c_{2}$, i.e.,
\be
c_{2} < {(n - 3)^2 \over 18} (n - 4) ~\mid \na \phi \mid^2~b^2.
\ee
\par
Let $p_{min}$ be a minimum of
$\psi$, then $\psi \circ \ga$ is a solution of the Cauchy problem for differential equation
\be
{d^2 \psi \over ds^2} = \bigg[
{3 c_{1} + 4(n - 3)\big[ (n - 3)(n - 6) + 2(n - 2) \big] \mid \na \phi \mid^2~\psi^3
\over 12 \psi^2} \bigg],
\ee
with the conditions $\psi(0) = p_{min}$ and ${d\psi \over ds}(0) = 0.$
As in Ref.\cite{chr05} we conclude that $\psi$ depends only on
geodesic distance from $ p_{min}$ and not on the direction of the geodesics.
This leads to the conclusion that the level sets of $\psi$ coincide with the geodesic sphere centered at 
$p_{min}$ (within the injectivity radius of $p_{min}$). The same considerations take place for the case of $p_{max}$,
where $p_{max}$ is the maximum of $\psi$ .
\par
On the set $\Omega = \{ {}^{(n-2)}\na \psi \neq 0 \}$, $\psi$ may be locally used as a coordinate. Thus, the metric
may be written in the form as
\be
{d \psi \over F^{2}(\psi)} + H^{2}(\psi, \xi) d\xi^2,
\ee
where $\xi$ is a local coordinate on the level set of $\psi$, while $F^{2}(\psi)$ is
given by
\be
F^{2}(\psi) = \mid {}^{(n-2)}\na \psi \mid^{2} = c_{2} - {c_{1} \over 2\psi} +
{1 \over 6}~ K~ \mid \na \phi \mid^2.
\ee
Inside the radius of injectivity of $p_{min}$, one has that ${d \psi \over F^{2}(\psi)} = d\rho$, where
$\rho$ is the distance function from $p_{min}$. Normalization of $\xi$ to run from zero to $2 \pi$
and by redefinition $\xi \rightarrow \la \xi$ provides
the condition that
$F^{2}(\psi) = \la^2 \rho^2 + \cO (\rho^2)$. 
This is sufficient to establish
a diffeomorphism among the level sets of 
$\psi$ within $\Omega$. 
\par
In the radius of injectivity of $p_{max}$ the arguments are the same, leading to the 
condition
$F^{2}(\psi) = \la^2 \hro^2 + \cO (\hro^2)$, where now $\hro$ is a distance function from $p_{max}$.
Eliminating $\la$ one concludes that the following implies
\be
3 c_{1} (b^2 + a^2) = - 2 (n - 3)^2~\mid \na \phi \mid^2~
a^2 b^2 (a + b).
\label{ca}
\ee
The other condition for $c_{1}$ we may obtain from the relation $F(a) = F(b) = 0$, i.e.,
\be
c_{1} = - {1 \over 3}(n - 3)^2~\mid \na \phi \mid^2~
a b (a + b).
\label{cb}
\ee
Substituting $c_{1}$ from (\ref{cb}) to relation (\ref{ca}) we get that $(a - b)^2 = 0$, i.e.,
$a = b$ which leads to a contradiction.
In the light of the above considerations, one can see that the regularity of the metric
requires the condition that $\mid {}^{(n-2)}\na \psi \mid = 0$. This provides the fact that
$\la = const$. It implies from Eq.(\ref{lam}) that $h_{(0)ab}$ is the metric of $S^{n-2}$ sphere.

\par
Our next step is to 
define on the hypersurface $\Sigma$ the orbit metric $\ga_{ab}$ defined as
follows:
\be
\ga(Y,Z) = g(Y, Z) - {g(X, Y)~g(X, Z) \over g(X, X)},
\ee
where by $X$ we have denoted the Killing vector which asymptotes ${\p \over \p t}$ in the asymptotic region.
It also fulfills the staticity condition. As in Ref.\cite{gib03}-\cite{gib02b} we shall consider conformal transformation
of the form $g_{\mu \nu} = \Omega_{\pm}^2 \ga_{\mu \nu}$, where
\be
\Omega_{\pm} = \bigg[ {(1 \pm V)^2 - \phi^2 \over 4} \bigg]^{1 \over n-3}.
\ee
As in our previous work \cite{rog03} $\phi$ satisfies the condition
\be
0 \leq \mid \phi \mid \leq 1 - V.
\label{ph}
\ee
One can see that the conformal factors $\Omega_{\pm}$ are nonnegative. The inequalities
are strict except when the metric is locally of Majumdar-Papapetrou type \cite{chr99b,rog03}. The possibility
that $\mid \phi \mid$ is strictly bounded away from one gives us the Reissner-Nordsr\"om solutions.\\
On the contrary,
we assume a contradiction. Namely, we suppose that there exists a component of the event horizon $\cH$ on which
$\phi = \pm 1$ and in this case $\cH$ is degenerate. Eq.(\ref{ph}) provides that
$\phi = \pm (1 - V + \cO(r ^{n-2}))$.
Taking into account the fact that $V = \mid g_{uu} \mid^{1\over 2} = \sqrt{A_{(0)}} r + \cO(r ^{n-2})$,
one gets the following expressions for the conformal factors:
\ben
\Omega_{+} &=& A_{(0)}^{\beta \over 2}~r^{\beta} + \cO(r^{n-2 + \beta}),\\
\Omega_{-} &=& \cO(r^{n-2 \over n-3}),
\een
where $\beta = 1 /(n-3)$.\\
Thus, it leads to the following form of the conformal metric
\ben
ds^2_{\pm} &=& { \Omega^{2}_{\pm}~ g_{uu} \over g_{uu}} \ga_{\mu \nu} =
A_{(0)}^{\beta-1}~r^{2 (\beta - 1)} \bigg[
1 + \cO(r^\beta) \bigg] \bigg[
dr^2 + \cO(r^2) dx^{a}dr \\ \nonumber
&+& r^2 (1 + \cO(r))A_{(0)}~h_{(0) ab} ~dx^{a}dx^{b} 
+ \cO(r^3)~dx^{a}dx^{b} \bigg].
\label{m1}
\een
The coordinate $r$ can be thought as a radial coordinate, As was shown $h_{(0)ab}$ is the metric
of the sphere, just $A_{(0)}h_{(0)ab}$ is the unit $S^{n-2}$ sphere. One has the following:
\be
r^2 A_{(0)}h_{(0)ab} dx^a dx^b = dx_{m}~dx^{m} - dr^2.
\label{met2}
\ee
This part of the metric in Eq.(\ref{met2}) combines with the leading part of the term $dr^2$ in relation (\ref{m1}).
The term $\cO(r^\beta)dr^2$ gives a contribution which in coordinates $x^{m}$ vanishes 
at the origin of coordinates as $\cO(x^i)$. On the other hand,
the similar reasoning as presented in Ref.\cite{chr05} shows that $\cO(r^2)dx^{a}dr$ gives the contribution 
in $x_{i}$ coordinates 
of the form
$r (g(x_{a}/r)) dx^{i} dx^{j}$, where $g(x_{a}/r))$ is a smooth function. The analysis of
$dx^{m} dx^{n}$ terms reveals that $ds^2_{\pm}$ can be rewritten as
\be
ds^{2}_{+} = A_{(0)}^{\beta-1}~r^{2 (\beta - 1)} \bigg(
\delta_{ij} + \cO(x^i) \bigg)~dx^{i} dx^{j}.
\label{dss}
\ee
One can see that in four-dimensional case we get the limit treated in Ref.\cite{chr05},
namely $ A_{(0)}^{\beta-1}~r^{2 (\beta - 1)} = 1$. 
\par
It is convenient to rescale $g_{\mu \nu (+)}$ in such a way that its scalar curvature vanishes.
Just we shall look for a
conformal transformation of the metric $g_{\mu \nu (+)}$ such that the scalar curvature 
vanishes. We construct $\psi^{2 \alpha}g_{\mu \nu (+)}$ to be the solution of the following relation:
\be
\psi^{2 \alpha + 1} R \bigg( \psi^{2 \alpha}g_{\mu \nu (+)} \bigg) 
= R \bigg( g_{\mu \nu (+)} \bigg) \psi
- 2 {}{}^{g_{(+)}} \na^{\mu}~{}{}^{g_{(+)}} \na_{\mu} \psi = 0,
\ee
where $\alpha = {1 \over n-1}$.\\
Then, one can search for $\psi$ in the form as $\psi = 1 + u$, where $u$ vanishes on 
a coordinate sphere of radius $b$.
It implies
\be
- 2 {}{}^{g_{(+)}} \na^{\mu}~{}{}^{g_{(+)}} \na_{\mu} u +  
R(g_{\mu \nu (+)})u = -  R(g_{\mu \nu (+)}).
\label{uuu}
\ee
For some constant the scalar curvature $R \bigg( g_{\mu \nu (+)} \bigg)$ satisfies the inequality
$R \bigg( g_{\mu \nu (+)} \bigg) \leq C/ \mid \vec x \mid$.
It can be seen from Eq.(\ref{dss}) and the fact that the first and the second derivative of
$g_{\mu \nu (+)}$ is bounded. Using the similar arguments as presented in Ref.\cite{chr06}
one can show the existence of a solution $u_{\ep}$ to the equation (\ref{uuu}) vanishing 
both on the coordinate sphere of radius one as well as on radius $\ep$.
\par
In the case of $g_{\mu \nu (-)}$ one can write the following:
\be
g_{\mu \nu (-)} = \bigg(
{\Omega_{-}^{2} \over \Omega_{+}^2} \psi^{-2 \alpha} \bigg)
\tg_{\mu \nu (+)} = \hpsi^{2 \alpha} \tg_{\mu \nu (+)},
\ee
where $\tg_{\mu \nu (+)} = \psi^{2 \alpha} g_{\mu \nu (+)}$.\\
It can be checked that under the conformal transformation the scalar curvature tensor has the form as
\be
\hpsi^{2 \alpha + 1} R \bigg( g_{\mu \nu (-)} \bigg) = R \bigg(\tg_{\mu \nu (+)} \bigg) \psi
- 2 {}{}^{\tg_{(+)}} \na^{\mu}~{}{}^{\tg_{(+)}} \na_{\mu} \hpsi.
\label{rrrrr}
\ee
The first term of the right-hand side of Eq.(\ref{rrrrr}) vanishes due to our previous construction.
Applying the maximum principle and the Serin removable singularity theorem \cite{ver} as well
as the comparison principle one gets that $\hpsi$ at the origin of coordinates $x^i$ is greater
than zero (see for the details Ref.\cite{chr05}). 
On the other hand, having in mind the form of conformal factors $\Omega_{\pm}$, one concludes
that this contradicts the fact that $\Omega_{-} / \Omega_{+} \rightarrow 0$ as we approach
the origin of the coordinates.
Then, one obtains that $\phi$ 
is equal $\pm 1$ at the origin of coordinates. This situation is only possible for
Majumdar-Papapetrou solutions. This fact leads as to the main conclusion of the work, namely:\\
\noindent
{\bf Theorem:}\\
Let us consider a static solution to the $n$-dimensional Einstein-Maxwell
equations of motion with an asymptotically timelike Killing vector field $k_{\mu}$.
Suppose, further that the manifold under consideration consists 
of a connected and simply connected
spacelike hypersurface $\Sigma$ to which $k_{\mu}$ is orthogonal. The topological
boundary $\p \Sigma$ of $\Sigma$ is a nonempty topological manifold with
$h_{ij}k^{i}k^{j} = 0$ on $\p \Sigma$.
Then, one arrives at the following conclusions:

\begin{enumerate}
\item{If $\p \Sigma$ is connected, then there exist a neighbourhood of $\Sigma$ which is isometrically
diffeomorphic to an open subset of $n$-dimensional Reissner-Nordstr\"om 
spacetime (extreme or non-extreme).}
\item{If $\p \Sigma$ is not connected then there is
an open neighbourhood of $\Sigma$ which is isometrically diffeomorphic to $n$-dimensional
Majumdar-Papapetrou spacetime.}
\end{enumerate}                      
 
\vspace{0.5cm}
\noindent
{\bf Acknowledgements:}\\
MR is grateful for hospitality to the Isaac Newton Institute, Cambridge, where the part of this work
was begun and P.Chru\'sciel for helpful discussion. The author  
was supported in part by the Polish Ministry of Science and Information
Society Technologies grant 1 P03B 049.



\end{document}